\documentclass[%
 reprint,
amsmath,amssymb,
aps,onecolumn,
pra,
floatfix,
justification=raggedleft]{revtex4-2}

\usepackage{graphicx} 
\usepackage{dcolumn} 
\usepackage{bm} 
\usepackage{comment}
\usepackage{cancel}
\usepackage{subcaption}

\usepackage{xcolor}

\begin{document}

\title{
Smarter Usage of Measurement Statistics Can Greatly Improve Continuous Variable Quantum Reservoir Computing
}

\author{Markku Hahto}
\affiliation{
 Department of Physics and Astronomy, University of Turku, FI-20014, Turun Yliopisto, Finland.
}

\author{Johannes Nokkala}
\affiliation{
 Department of Physics and Astronomy, University of Turku, FI-20014, Turun Yliopisto, Finland
}
\affiliation{
 Turku Collegium for Science, Medicine and Technology, University of Turku, Turku, Finland.
}

\date{July 4, 2025}

\begin{abstract}
Quantum reservoir computing is a machine learning scheme in which a quantum system is used to perform information processing. A prospective approach to its physical realization is a photonic platform in which continuous variable (CV) quantum information methods are applied. The simplest CV quantum states are Gaussian states, which can be efficiently simulated classically. As such, they provide a benchmark for the level of performance that non-Gaussian states should surpass in order to give a quantum advantage. 
In this article we propose two methods to extract more performance from Gaussian states compared to previous protocols. We consider better utilization of the measurement distribution by sampling its cumulative distribution function. We show it provides memory in areas that conventional approaches are lacking, as well as improving the overall processing capacity of the reservoir. We also consider storing past measurement results in classical memory, and show that it improves the memory capacity and can be used to mitigate the effects of statistical noise due to finite measurement ensemble.
\end{abstract}

\maketitle


\section{Introduction} \label{sec:introduction}

Quantum reservoir computing (QRC) is a quantum machine learning paradigm in which the dynamics of a generic quantum system are harnessed for information processing \cite{mujal2021opportunities}. Unlike in the more common models for neural networks where the weights of connections in the network are trained to optimize performance, in reservoir computing the system playing the role of the network is fixed. The training only happens at a classical linear readout layer, which gets its inputs from measurements of the reservoir state. When utilizing quantum systems as the reservoir, the exponentially growing Hilbert space dimension could in principle allow a handful of qubits to compete with hundreds of classical neurons \cite{fujii2017harnessing}. All in all, quantum reservoir computers boast a unique combination of amenability to custom hardware implementations, extremely rapid and robust training, and high computational power. These favorable features have led QRC to enjoy a steadily growing interest as a candidate for a future quantum technology \cite{ghosh2020reconstructing,senanian2024microwave,krisnanda2025experimental}.

QRC generally faces a few significant obstacles \cite{mujal2021opportunities}: the measurement of the systems collapses their state, leading to loss of information about the past state of the reservoir. Weak measurements \cite{mujal2023time} or feeding the measured values back to the reservoir \cite{kobayashi2024feedback,monomi2025feedback,paparelle2025experimental} have been proposed to alleviate this issue. Similarly, decoherence-induced noise and loss of memory are problems particularly in noisy intermediate scale quantum (NISQ) era qubit based approaches. Continuous-variable QRC implemented in quantum optical platforms presents a promising alternative: its extreme robustness to decoherence gives it improved scalability, and it can naturally facilitate coherent feedback while evading measurement back-action \cite{garcia2023scalable}. As drawbacks, current continuous-variable platforms are mostly limited to Gaussian resources which have limited nonlinearity \cite{nokkala2021gaussian,nokkala2022high} and only a quadratic advantage over comparable classical alternatives \cite{nokkala2021gaussian}. Additionally, QRC in general is limited by statistical noise arising from finite ensemble effects.

Here we address the main weaknesses of continuous-variable QRC by introducing a novel computational resource which is experimentally convenient and unique to CV platforms. We show how even with only Gaussian resources it leads to a vastly increased computational power and lifts the limitations on nonlinearity. We also show how to alleviate the detrimental effects of finite ensemble effects by using past measurement results stored in classical memory as additional output features. Our results push the limits of what can be easily achieved with just Gaussian resources, establish a new benchmark that any non-Gaussian generalizations should beat, and are readily applicable beyond Gaussian states.

Specifically, we go beyond the conventionally used moments of the measurement outcome distribution. Many schemes rely on means and covariances \cite{nokkala2021gaussian,nokkala2022high,garcia2023scalable,nokkala2024retrieving} which indeed completely determine a Gaussian state. Using higher order moments has been proposed to introduce the nonlinearity missing from the lower moments \cite{garcia2024squeezing}. This amounts to introducing nonlinearity in the classical post-processing phase. We will instead use the cumulative distribution function (CDF), which is sampled by simply tallying the proportion of the measurement outcomes below a given threshold value. Another key strength of using the CDF is that extracting more computational power out of a single observable can now be achieved by simply changing the threshold value rather than computing nonlinear functions of increasing degree. Importantly, for discrete variables CDF is linear in the probabilities of individual measurement outcomes and consequently does not constitute a source of additional computational power. 

The use of classical memory as extra computational nodes has been studied in classical reservoir computing literature to reduce the size of the reservoir while retaining computational capabilities \cite{sakemi2020model,duan2023embedding}, whereas here we introduce it for 
the purpose of reducing the size of the measurement ensemble needed to extract meaningful information from the quantum reservoir.
It can be readily used in conjunction with previously introduced means such as utilizing squeezing \cite{garcia2024squeezing}, or applying standard signal processing methods \cite{ahmed2025optimal}.

The rest of this article is structured as follows. Section~\ref{sec:methods} introduces the scheme for photonic reservoir computing and Gaussian states. In Section~\ref{sec:results} we present our results. We conclude with discussion about the results in Section~\ref{sec:conclusion}.


\section{Methods} \label{sec:methods}

\subsection{Quantum reservoir computing}

\subsubsection{Photonic platform}

\begin{figure}
        \centering
        \includegraphics[width=0.95\linewidth]{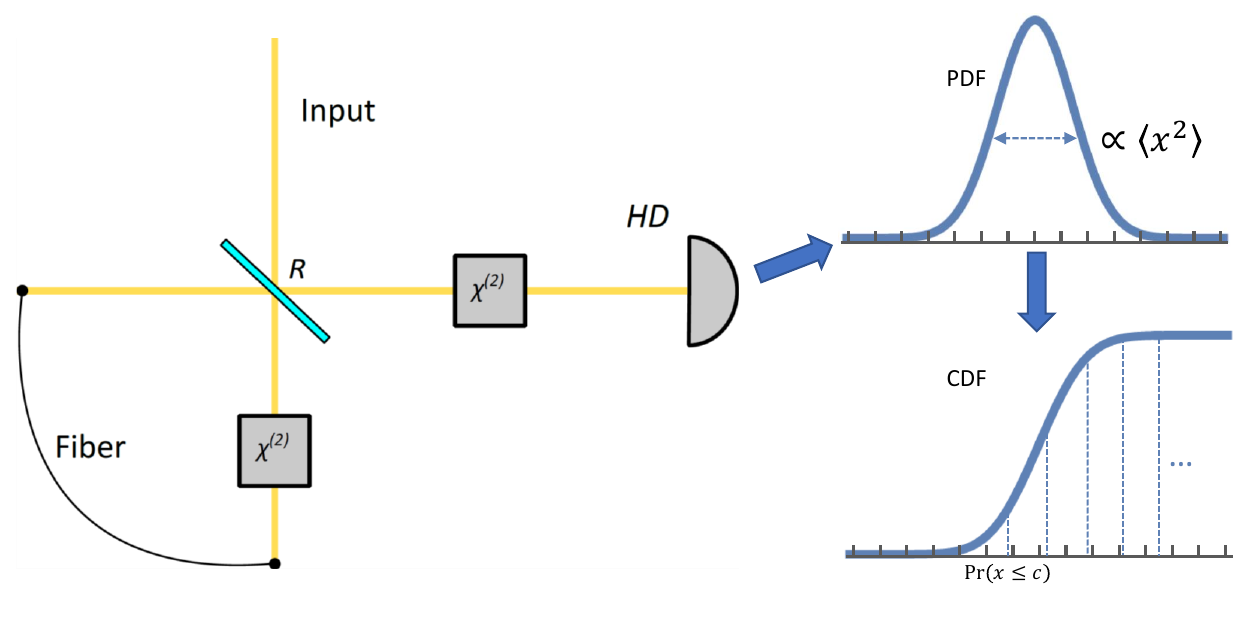}
        \caption{Schematic of the photonic reservoir computer and its use. The reservoir is a physical ensemble of optical modes circulating in a fiber loop. The classical input modulates squeezed vacua which are injected through a beam splitter, which is also responsible for output extraction. $\chi^{(2)}$-crystals are responsible for creating correlations. Conventionally, homodyne measurements of the $x$-quadratures of each mode are used to estimate the expected variance $\langle x^2\rangle$ of the normally distributed measurement outcomes which become the computational nodes. Here we introduce the use of the cumulative distribution function (CDF) to form the nodes. It can be sampled for each measured observable at various points $c$ by evaluating the probability that a measurement outcome is at most $c$.
        In both cases tasks are accomplished as is typically done, i.e. by forming the optimal linear combination of the computational nodes.}
        \label{fig:qrc_schematic}
\end{figure}

We adopt the photonic platform introduced in Ref.~\cite{garcia2023scalable}. In this setup, the reservoir consists of a pulse of $N$ optical modes traveling in a loop of optical fiber, coupled with a nonlinear $\chi^{(2)}$ crystal. The inputs are encoded into $N$ ancillary modes, which couple to the reservoir modes through a beam splitter with reflectivity $R$. One output of the beam splitter is sent to a detector through another nonlinear crystal, and the other is fed back to the loop of fiber to keep a recurrent pulse traveling in the loop. To create an ensemble of reservoirs, the input pulses are repeated $M$ times. Figure~\ref{fig:qrc_schematic} shows a schematic of the platform. In this article we fix the number of modes to $N = 7$ and the reflectivity of the beam splitter to $R = 0.75$.

The Hamiltonian for the $\chi^{(2)}$ crystal is
\begin{equation}
    H = \sum_{j=1}^{N} \omega_j \left( {a}^\dagger_j{a}_j + \frac{1}{2}  \right) + \sum_{j>k} \left( g_{jk}{a}^\dagger_j{a}_k + i h_{jk}{a}_j^\dagger{a}^\dagger_k + \text{h.c.}  \right), \label{eq:nonlinear_ham}
\end{equation}
where $g_{jk}$ and $h_{jk}$ are random real constants defining the couplings between modes in the crystal, and  ${a}^\dagger_j$ and ${a}_j$ are the creation and annihilation operators for each mode. We set $g \in [0.1,0.3]$ and $h \in [0.2,0.4]$ following Ref.~\cite{garcia2023scalable}. 

We restrict the discussion to zero mean Gaussian states. They are fully described by their covariance matrix $(\sigma(\mathbf{x}))_{kl} = \langle \mathbf{x}_k \mathbf{x}_l + \mathbf{x}_l \mathbf{x}_k \rangle/2 - \langle\mathbf{x}_k \rangle\langle\mathbf{x}_l \rangle$, where $\mathbf{x}= \{x_1, p_1, x_2, p_2,\dots\}^\top$ is the vector of quadrature operators of each mode. The quadrature operators are defined as ${x} = {a}^\dagger + {a}$ and ${p} = i({a}^\dagger - {a})$. As such, the covariance matrix of the vacuum is equal to the identity. Going forward, we drop the dependence from $\mathbf{x}$ from the notation and denote $\sigma \equiv \sigma(\mathbf{x})$. Since we only consider Gaussian initial states and Hamiltonians which are quadratic in the creation and annihilation operators, the states remain zero mean and Gaussian throughout \cite{weedbrook2012gaussian}. 

On each time step $k$, the input $s_k$ is encoded into the $N$ ancillary modes as a product state of Gaussian states $(\sigma_k^{\text{in}})$ of the form in Eq.~\ref{eq:covariance}. We consider encoding in the squeezing magnitude $r$, such that $r_k = (s_k+1)/2$ which leads to $r_k \in [0,1]$. The total covariance matrix is $(\sigma_k^{\text{in}})^{\oplus N}$, from which we drop the exponent for clarity. The reservoir's state has the covariance matrix $\sigma_{k-1}^{\text{R}}$. Thus the initial state of the combined system of the ancillae and the reservoir is $\sigma_k^{\text{(R + in)}} = \sigma_{k-1}^{{\text{R}}}\oplus\sigma_k^{\text{in}}$ because there are no correlations between the reservoir and the input modes. The action of the beam splitter and the two nonlinear crystals on the covariance matrix can be represented by a symplectic matrix $S(\Delta t)$, with which the state at the end of the step is 
\begin{equation}
    \sigma_k^{\text{(R + out)}} = S(\Delta t) \sigma_k^{\text{(R + in)}} S(\Delta t)^\top    \label{eq:cov_evolution}
\end{equation}
The reservoir dynamics create correlations between the reservoir and the output modes, and thus the covariance matrix $\sigma_k^{\text{(R + out)}}$ has the block matrix form
\begin{equation}
    \sigma_k^{\text{(R + out)}} = \begin{pmatrix}
       \sigma_k^{\text{R}}  & \sigma_k^{\text{corr}} \\ 
        \sigma_k^{\text{corr}}& \sigma_k^{\text{out}} 
    \end{pmatrix}.
\end{equation}
The reservoir state is updated to $\sigma_k^{\text{R}}$. Details of Gaussian states and the symplectic formalism are explained in App.~\ref{app:gaussian}.

Meanwhile, the output modes are sent to a homodyne detector which measures the $x$-quadratures of $\sigma_k^{\text{out}}$. This amounts to extracting the corresponding $N \times N$ submatrix from the covariance matrix. In Ref.~\cite{garcia2023scalable} it was shown that the measurement back action on the reservoir state averages out and can be neglected. Due to the symmetry of the covariance matrix, the measurements yield a total of $N(N+1)/2$ unique values to form the output vector $\mathbf{o}$ of the reservoir.

In the ideal case of an infinite number of measurements, the covariances in $\sigma_k^{\text{out}}$ can be measured exactly. In reality, however, there is a finite number of samples $M$ in the ensemble, from which a maximum likelihood estimate for $\sigma_k^{\text{out}}$ must be calculated. This estimate, $\tilde{\sigma}_k^{\text{out}}$, can be modeled by the Wishart distribution $W(\sigma_k^{\text{out}},M)$ \cite{wishart1928generalised}, such that 
\begin{equation}
    \tilde{\sigma}_k^{\text{out}} = \frac{1}{M}W(\sigma_k^{\text{out}},M).\label{eq:wishart}
\end{equation}
In the limit $M \to \infty $, the ideal covariance matrix $\sigma_k^{\text{out}}$ is recovered.

\subsubsection{Reservoir computing}

Reservoir computers are especially suitable for time series processing, where tasks consist of an input sequence $\mathbf{s}$ and a target sequence $\bar{\mathbf{y}}$. The reservoir processes the input such that $\mathbf{O} = f(\mathbf{s})$, where $f$ is a function mapping the input sequence $\mathbf{s}$ to $\mathbf{O}$, which is the output matrix of reservoir observables. The output matrix can be written as $\mathbf{O}= \left( \mathbf{o}_1, \,\mathbf{o}_2, \, \mathbf{o}_3,\dots\right)^\top$  where $\mathbf{o}_i$ are vectors of the reservoir observables of each time step. 

The sequence $\mathbf{s}$ is divided into three parts: preparation, training and testing. The preparation period (also called wash-out) guarantees that the echo state property is fulfilled for the training and testing data, i.e., they are independent of the initial state of the reservoir. During the training and testing periods the reservoir observables are collected to two matrices $\mathbf{O}_{\text{train}}$ and $\mathbf{O}_{\text{test}}$. Training the reservoir computer is done by calculating the weight matrix $\mathbf{W}$ which minimizes the squared error between the prediction $\mathbf{y}_{\text{train}} = \mathbf{W}\mathbf{O}_{\text{train}}$ and the target $\bar{\mathbf{y}}_{\text{train}}$. The optimal weights are easily found with
\begin{equation}
    \mathbf{W} = \mathbf{O}_{\text{train}}^+ \bar{\mathbf{y}}_{\text{train}}^\top
\end{equation}
where $\mathbf{O}_{\text{train}}^+$ is the Moore-Penrose pseudoinverse of $\mathbf{O}_{\text{train}}$. 

Performance of the reservoir is then evaluated by comparing the predicted outputs in the testing phase $\mathbf{y}_{\text{test}} = \mathbf{W}\mathbf{O}_{\text{test}}$ to the target $\bar{\mathbf{y}}_{\text{test}}$. As the figure of merit we use normalized mean square error (NMSE), defined as
\begin{equation}
    \text{NMSE}(\bar{\mathbf{y}},\mathbf{y}) = \frac{\sum (\bar{\mathbf{y}}-\mathbf{y})^2}{\sum \bar{\mathbf{y}}^2}, \label{eq:nmse}
\end{equation}
so that the reservoirs ability to recreate the target sequence is given by $\text{NMSE}(\bar{\mathbf{y}}_{\text{test}},\mathbf{y}_{\text{test}})$. Further, the capacity of a reservoir on a task is defined as
\begin{equation}
    C_{\bar{\mathbf{y}}} = \max[0, 1-\text{NMSE}(\bar{\mathbf{y}},\mathbf{y})]. \label{eq:capacity}
\end{equation}
When the reservoir perfectly replicates the target sequence, $C_{\bar{\mathbf{y}}} = 1$, and when the output and target have no correlation, $C_{\bar{\mathbf{y}}}= 0$.


\subsection{Tasks}

We consider some standard figures used in the reservoir computing literature to quantify the information processing capacity of the reservoir. As an example of a task which is difficult to handle with the chosen reservoir scheme, we use nonlinear autoregressive moving average (NARMA) \cite{kubota2021unifying}. To compare the performance across different tasks we employ information processing capacity (IPC) \cite{dambre2012information}. For each task the input sequence $\mathbf{s}$ consists of random numbers drawn uniformly from $[-1,1]$.

NARMA\textit{n} (nonlinear auto-regressive moving average) is a nonlinear task where the reservoir has to process cross terms of inputs at different delays. The target $\bar{y}$, with input sequence $\mathbf{s}=\{s_k\}$, is given by
\begin{equation}
    \bar{y}_{k+1} = \alpha \bar{y}_k + \beta\sum_{\tau=0}^{n-1} \bar{y}_{k-\tau} +\gamma s_{k-n+1} s_k + \delta,    \label{eq:narma}
\end{equation}
with the parameters fixed as $\alpha = 0.3$, $\beta = 0.05$, $\gamma = 0.0375$ and $\delta=0$, following Ref.~\cite{nokkala2022high}. It consists of the moving average of the past $n$ output values, and an auto-regression term of the product of the input at time $k$ and time $k-n+1$. 
The output covariance matrices have been shown to lack cross terms of inputs at different times, and thus any NARMA\textit{n} task is impossible to solve with the standard approach of CV-QRC \cite{nokkala2022high}.

Performance in NARMA\textit{n} is a showcase of the cross terms and nonlinear processing. For a more precise quantification of a reservoir's capacity we employ the information processing capacity. IPC quantifies the processing done by the reservoir by testing its ability to recall past inputs, and powers and products thereof. It is defined as
\begin{equation}
    \text{IPC} = \sum_{\bar{\mathbf{y}}\in \{\bar{\mathbf{y}}\}} C_{\bar{\mathbf{y}}}, \label{eq:IPC}
\end{equation}
where $\{\bar{\mathbf{y}}\}$ is chosen to be a complete set of orthogonal functions \cite{dambre2012information}. One such set is formed by taking products of normalized Legendre polynomials
\begin{equation}
    \bar{y}_{\{d_i\}} = \prod_{i} \mathcal{P}_{d_i}(s_{k-i}),
\end{equation}
where $\mathcal{P}_{d_i}$ is the Legendre polynomial of degree $d_i$, and $s_{k-i}$ is the input delayed by $i$. The sequences are labeled by ${\{d_i\}}$, which denotes which degree polynomial is taken of which delayed input. Some details on the calculation of the IPC are included in App.~\ref{app:IPC}.

Theoretically, computing the IPC requires an infinite input sequence. However, in practice this is infeasible due to finite available computing time. We fix the length of the input sequence to be $10500$ steps, which is divided into wash-out ($500$), training ($8000$) and testing ($2000$). This wash-out period is sufficiently long to remove the influence of the initial state of the reservoir with the chosen reservoir parameters \cite{garcia2023scalable}. 


\section{Results} \label{sec:results}

When the observables of the reservoir are measured, a distribution of measurement results is received. For Gaussian states the distribution is completely defined by its first two moments, covariance and mean (or covariance matrix and vector of means in the multimode case). Therefore the reservoir outputs are often reduced to just these two quantities \cite{nokkala2021gaussian}. Higher order moments can also be used \cite{garcia2023scalable,garcia2024squeezing}, but for Gaussian states they are merely polynomial in the covariances and thus provide limited nonlinearity.  

While the Gaussian distribution is defined by the moments, other measurement statistics can also be considered. Here we show that the experimentally simple statistic of the cumulative distribution function (CDF) provides a high degree of nonlinearity even for Gaussian states. 
Let us consider the measurement distribution for the $x$-quadrature of a Gaussian state.
The CDF of the distribution is defined as the probability that a measurement $X$ falls below a given value $c$, i.e., $F_X(c) = \mathrm{Pr}(X \leq c)$. For measured data this corresponds to calculating the fraction of observations that fall below $c$, making the CDF simple to calculate from experimental data.
Analytically the CDF of a Gaussian distribution is slightly more complicated: while it is completely determined by the mean $\langle X\rangle$ and variance $\langle X^2\rangle$, it has no closed form expression. However, it can be expanded into a series, such that
\begin{equation}
    F_x(c) = \frac{1}{2} + \frac{1}{\sqrt{\pi}}\left( z - \frac{z^3}{3} + \frac{z^5}{10} - \frac{z^7}{42} + \dots\right), \label{eq:cdf_taylor}
\end{equation}
with $z = \frac{c-\langle X \rangle}{\sqrt{2\langle X^2 \rangle}}$. It is then clear that the CDF is nonlinear in the variances.
Taking into account that the covariances at each time step are of the form $\langle X^2 \rangle = \sum_{\tau \geq 0} f_\tau(s_{m-\tau})$, with $f_\tau$ being functions of the past inputs, it can be seen that the CDF contains cross terms of inputs at different delays. It should also be noted that the nonlinearity and cross terms in the CDF are independent of chosen input encoding, in contrast with the covariance matrix approach \cite{nokkala2022high}. Some more details about the CDF for Gaussian distributions are found in App.~\ref{app:cdf}.

Since the output states of the reservoir are multimode Gaussian states, it is natural to also consider the multivariate CDF. It is defined as $F_{\mathbf{X}}(\mathbf{c}) = \mathrm{Pr}(\mathbf{X}\leq \mathbf{c}) = P(X_1 \leq c_1; X_2 \leq c_2;\dots)$, with $X_i$ being the measurement results of the individual modes. Although the multivariate CDF can be formally written as an integral of the joint probability density, for a Gaussian distribution the integral cannot be solved analytically and the CDF cannot be expressed as elementary functions. 
 
In addition to being nonlinear in the covariance, the values of the CDF at two different points $F_X(a)$ and $F_X(b)$ are not linearly dependent. The CDF can thus be sampled at multiple points to increase the number of outputs from the reservoir. This leads us to using the CDF as an output of the reservoir: first, we fix a set of points $\{c_a\}_{a=0}^{n_{UV}}$ at which the CDF is to be sampled. For each observable's measurement distribution at each time step, the CDF $F_{X_i}(c_a)$ is then sampled at these points. This yields $N \times n_{UV}$ outputs in addition to the $N(N+1)/2$ outputs from the covariance matrix. Similar procedure applies when multivariate CDFs are considered: the function $F_{\mathbf{X}}(\mathbf{c})$ is sampled at points $\{\mathbf{c}_a\}_{a=0}^{n_{MV}}$. For example, for a bivariate CDF the set $\{\mathbf{c}_a\}$ can be chosen to be a grid of points $[c_1, c_2,\dots] \times [c_1, c_2,\dots]$. The bivariate CDF yields $N (N-1)/2 \times n_{MV}$ outputs since the CDF of each pair of modes can be separately sampled.

Multivariate CDFs of more than two observables could also be considered. However, since for an $m$-variate CDF the number of outputs scales as $N^m$, the numerical calculations become significantly more time consuming: even with only bivariate CDFs our simulation runs some orders of magnitude slower than with univariate CDFs. Since the bivariate case already proves that this approach gives a boost in performance, we do not study higher multivariate CDFs.

In an experiment calculating the values of the CDF simply amounts to evaluating the fraction of observations for which $X_i \leq c_a$ (or $\mathbf{X} \leq \mathbf{c}_a$). However, since the simulations involve only the covariance matrices, the CDFs must be numerically calculated from the covariances. In the univariate case we simply take the diagonal entries of the $N\times N$ covariance matrices, while for the bivariate CDF we extract all $2 \times 2$ submatrices which are the covariance matrices for each pair of modes. With these (co-)variances, the values of the CDFs are calculated using the SciPy-library for the Python programming language \cite{virtanen2020scipy}.

Another approach to improving the performance of the reservoir, inspired by the consideration that classical processing and data storage is cheap compared to operating the quantum reservoir, is to utilize past measurement results in the predictions. Instead of forming the matrix of observables $\mathbf{O}$ from the outputs at each individual time step $k$ such that the rows are of the form $o_0 \oplus \mathbf{o}_k$, some number $P$ of past observations are used. The matrix $\mathbf{O}$ now has rows of $o_0 \oplus \mathbf{o}_k \oplus \mathbf{o}_{k-1} \oplus \dots \oplus \mathbf{o}_{k-P}$. As long as sequential measurements are linearly independent of each other, each included step should increase the capacity. Due to the nonlinear $\chi^{(2)}$-crystal in the setup and the measurements only extracting the $x$-quadratures---i.e., only a quarter of the total covariance matrix---it should be expected that the sequential measurements are not linearly dependent. This approach has been used in classical reservoir computing to decrease the size of the reservoir while retaining similar levels of performance as larger reservoirs \cite{sakemi2020model,duan2023embedding}. In the quantum context it can similarly help against the statistical noise from the finite ensemble size, effectively increasing the size of the ensemble.

In the numerical results for the following section section, we take for the univariate CDF $n_{UV}=10$ uniformly spaced points from the interval $[0.1, 2.0]$, and for the bivariate CDF we consider a uniform grid of points $[0.1,0.6,1.1,\dots]\times[0.1,0.6,1.1,\dots]$ for a total of $n_{BV} = 4, 9, 16$ points. We have checked that the actual values are less important than the number of sampling points, and for example sampling the univariate CDF at points from $[1.1, 3.0]$ results in similar performance as the chosen interval. However, if a limited number of samples is considered, care must be taken to choose the points so that they can be distinguished; if for example for some distribution $F_x(2) = 0.998$ and $F_x(2.2) = 0.999$, with less than 1000 samples these cannot be reliably distinguished.


\subsection{Processing capacities in the ideal case}

\begin{figure}[t]
    \centering
    \begin{subfigure}[t]{0.49\textwidth}
        \centering
        \includegraphics[width=\textwidth]{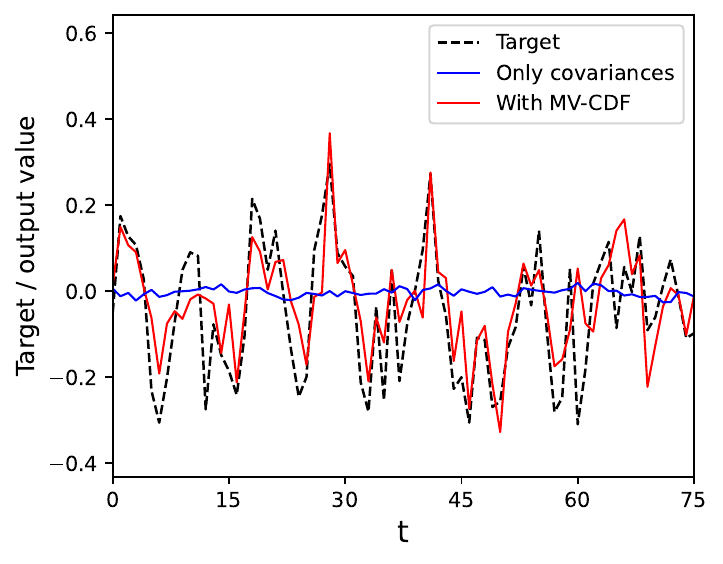}
        \caption{An example realization of NARMA10 task, showing the target, and predictions with only covariance matrices and with values from the bivariate CDF. When using only covariances the reservoir fails to recreate the target completely, whereas with the multivariate CDF the predictions follow the target quite closely. NARMA10 includes the cross term $s_k s_{k-9}$, which is absent from the covariances and evidently present in the CDFs.}
        \label{fig:cdf-narma-example}
    \end{subfigure}
    \begin{subfigure}[t]{0.49\textwidth}
        \centering
        \includegraphics[width=\textwidth]{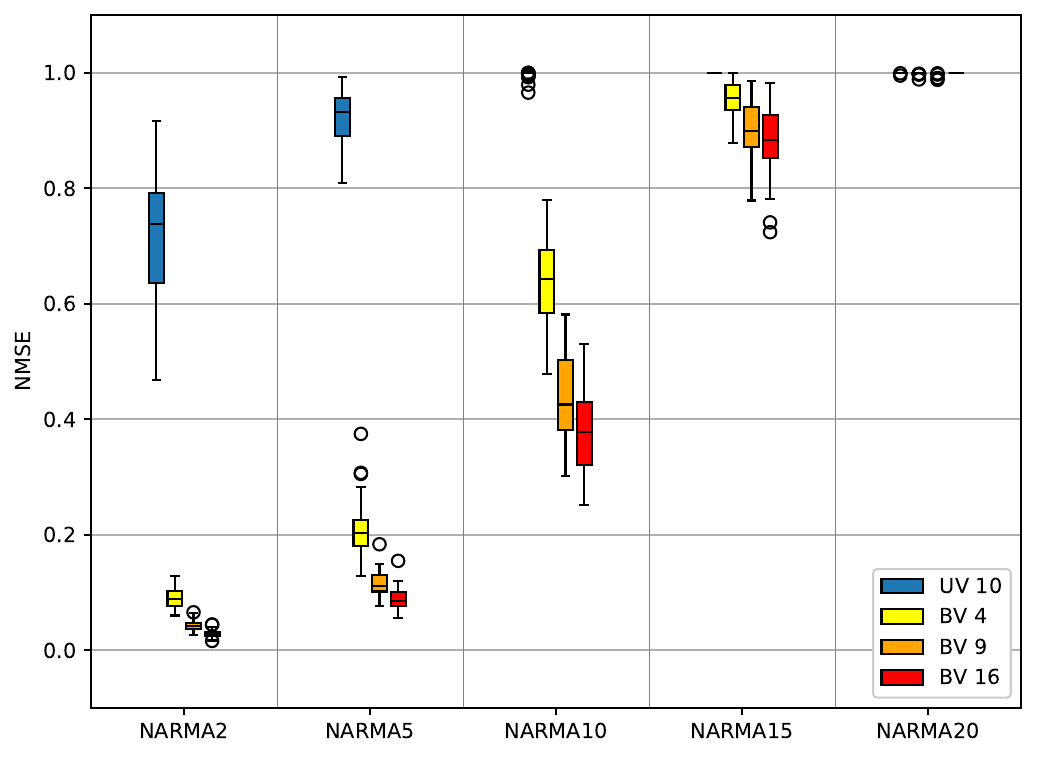}
        \caption{NMSE achieved in different NARMA$n$ tasks using different number of points from the uni- and bivariate CDFs (UV and BV). The tasks were performed with 50 realizations of the reservoir.}
        \label{fig:cdf-narma-nmses}
    \end{subfigure}
    \caption{Performance in the NARMA task using the CDF.}\label{fig:cdf-narma}
\end{figure}

Here we focus on the theoretical performance limits attainable in the case of infinite number of measurements, $M \to \infty$, where the observables have zero statistical noise and can be measured exactly. As discussed above, CDF contains cross terms of inputs at different delays, which lack from the covariance matrix approach. Therefore, using CDF should improve performance in the NARMA task. An example realization of the NARMA task is shown in Figure~\ref{fig:cdf-narma-example}, along with predictions using covariances and using bivariate CDF. It is clearly visible that the covariance matrices cannot predict the behavior at all but with the CDF the predictions match the target relatively well.

This point is expanded upon in Figure~\ref{fig:cdf-narma-nmses}, in which NMSEs in the NARMA$n$ task with different $n$ are presented. The covariance matrices fail to predict even NARMA2, thus having $\text{NMSE}=1$, while sampling the bivariate CDF results in almost perfect predictions with lower values of $n$ and non-vanishing prediction success with $n$ up to $15$. Figure~\ref{fig:cdf-narma-nmses} also highlights the fact that sampling more points from the CDF improves the performance, although with diminishing returns. Another feature of note is that while the NARMA task requires quadratic cross terms, the univariate CDF, as seen in Eq.~\eqref{eq:cdf_taylor}, only provides terms of odd degrees, which leads to relatively unimpressive performance.

\begin{figure}
    \centering
    \includegraphics[width=0.5\linewidth]{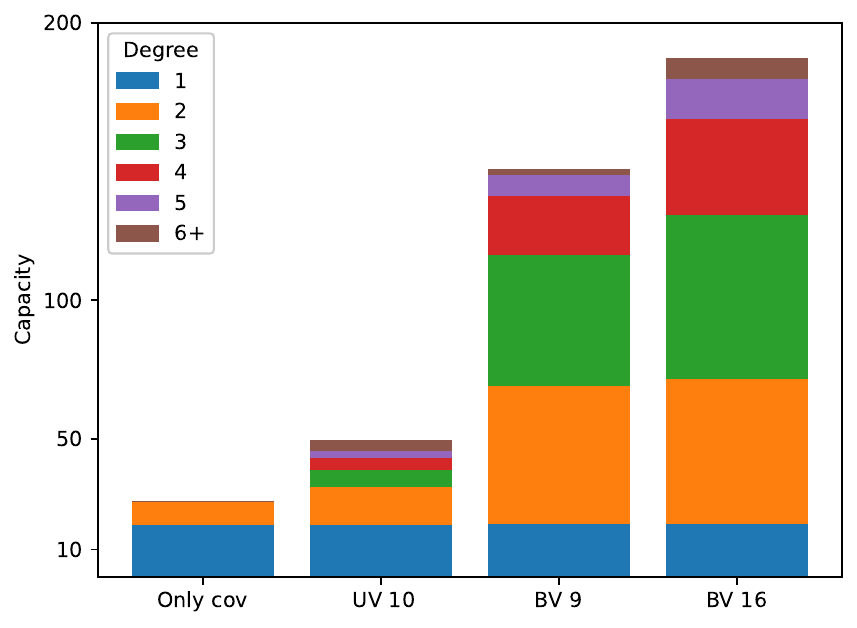}
    \caption{Information processing capacity of the CDF schemes by degree. Compared to the covariances-only approach the CDF schemes lead to substantially higher nonlinearity in terms of the degree reached as well as higher overall capacity.} Averaged over 50 realizations. 
    \label{fig:cdf-ipc-degree}
\end{figure}

While success in the NARMA task confirms that there are cross terms in the output, to fully quantify the performance gains from the CDF we study the the information processing capacities. Figure~\ref{fig:cdf-ipc-degree} shows the IPC separated by degree for the schemes. For all schemes the linear capacity remains the same. Sampling from the univariate CDF adds some nonlinear capacity to the reservoir, and the bivariate CDF increases the nonlinear capacity significantly. Sampling more points from the bivariate CDF particularly improves the higher degree capacities. It should be noted that while the theoretical upper bound for IPC is equal to the number of computational nodes, the CDF schemes do not reach this bound in our calculations. For example, with 9 samples from the bivariate CDFs of $N=7$ modes, the number of computational nodes is $217$, while from Figure~\ref{fig:cdf-ipc-degree} we see that $\mathrm{IPC}_{BV9}\approx 170$. This underestimation is to be expected, as explained in Appendix~\ref{app:IPC}.

\begin{figure}
    \centering
    \includegraphics[width=0.5\linewidth]{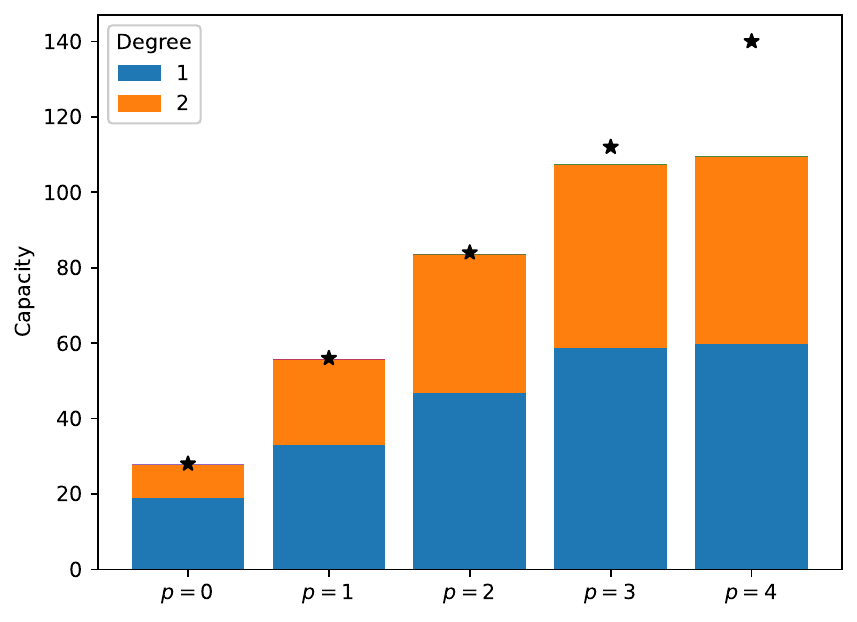}
    \caption{Information processing capacity of the classical memory schemes by degree. The black stars indicate the number of computational nodes, which corresponds to the theoretical upper bound on the IPC. Averaged over 50 realizations.}
    \label{fig:cm-ipc-degree}
\end{figure}

Similarly the effect of using past observations on the IPC of the reservoir may be quantified. In Fig.~\ref{fig:cm-ipc-degree} the IPC of the classical memory augmented scheme with only covariance matrices is presented up to $P = 4$. The processing capacity of the covariance matrices is limited to only degree 2 tasks with the chosen encoding. We see that with $P \leq 3$ the IPC bound is nearly achieved, while from $P = 3$ to $P = 4$ the improvement in capacity is marginal. This can be explained by looking at the rank of the output matrix $\mathbf{O}$: for $P \leq 3$ the rank is equal or almost equal to the number of columns, meaning that the output features are linearly independent. When $P$ increases to $4$, the rank barely increases, meaning that almost all information in the outputs of step $k-4$ is contained in the outputs of the later steps.

\subsection{Finite ensemble}

The results of the previous section consider the limiting case of $M \to \infty$ measurements, which is not experimentally realistic. Here we move to the more realistic scenario of a finite measurement ensemble and consider how the statistical noise from a finite number of measurements affects the capacity of the reservoir. 

\begin{figure}
    \centering
    \includegraphics[width=0.5\linewidth]{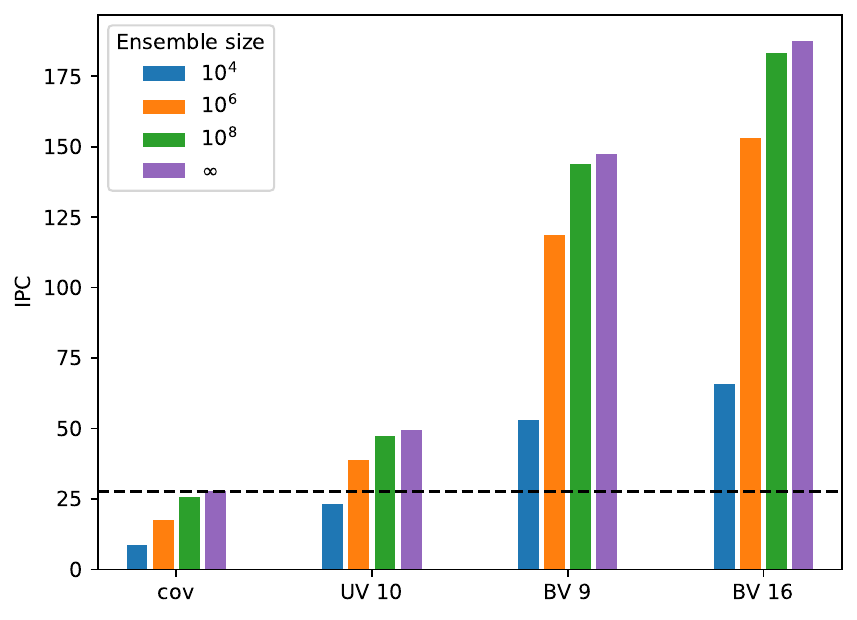}
    \caption{Information processing capacity of the different schemes with finite ensembles of various sizes and the ideal case of an infinite ensemble. The dashed black line shows the covariances-only performance when $M \to \infty$. Averaged over 50 realizations.}
    \label{fig:cdf-ipc-noisy}
\end{figure}

In Figure~\ref{fig:cdf-ipc-noisy} the effect of a finite ensemble is considered by sampling the covariance matrices according to Eq.~\eqref{eq:wishart}. The performance drops as the ensemble gets smaller, as expected. Comparing the loss of performance in the "covariances only"-approach to the CDF-based approaches, the scaling looks to be quite similar, although less severe; with only covariance matrices the ratio of IPCs for $M=10^6$ and the ideal case is around $0.6$, while for the bivariate CDF the ratio is around $0.8$. It appears that the CDF is slightly more resistant towards statistical noise than the covariance matrices. Another notable feature is that even with such a limited number of measurements as $10^4$, the IPC of the bivariate CDF method surpasses the ideal case limit for only covariances.

\begin{figure}
    \centering
    \includegraphics[width=0.5\linewidth]{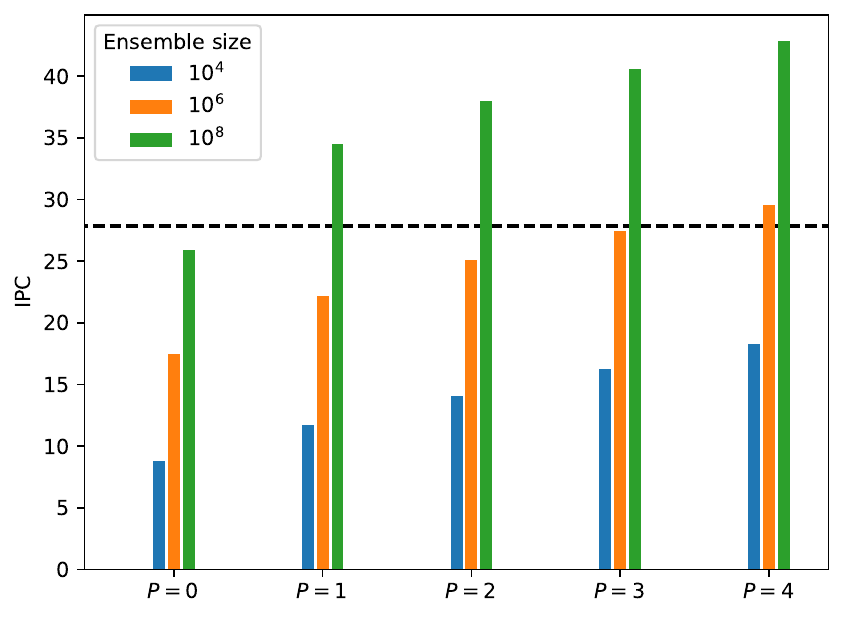}
    \caption{Using observables of past time steps can allow finite measurement ensembles to surpass the performance of the ideal case. The dashed black line shows the $P = 0$ performance when $M \to \infty$.
    Averaged over 50 realizations.}
    \label{fig:cm-ipc-noisy}
\end{figure}

We may also compare how the classical memory augmented scheme compares with finite ensembles. Fig.~\ref{fig:cm-ipc-noisy} shows the IPC of various ensemble sizes and values of $P$ compared to the ideal $P = 0$ performance. Again we observe that the IPC of the $P=4$ scheme beats the ideal performance of $P=0$ with $M=10^6$, and with a larger ensemble the requirement for $P$ decreases significantly. Compared to the ideal scenario we do not observe a similar decrease in performance gains going from $P=3$ to $P=4$. This is due to the uncorrelated noise between sequential time steps, which erases any linear dependence between sequential outputs.

\section{Conclusions and discussion} \label{sec:conclusion}

In this work we studied quantum reservoir computing with Gaussian states. Our results show that even though the measurement distribution of Gaussian states is completely determined by their means and variances, other statistical measures such as the cumulative distribution function can be used to easily extract additional information from the measurements, leading to significantly higher nonlinear processing capacity for the reservoir. We also have shown that storing past measurement results in classical memory and using them as additional features can be beneficial in mitigating the effect of a limited ensemble size, potentially even surpassing the ideal performance of the scheme without classical memory. Depending on the ensemble size, this can be achieved even by using just the previous time step.

More concretely, in the ideal case of infinite ensemble size $M=\infty$ we observed up to an almost ten-fold increase in total capacity when using the bivariate CDF and also significant improvements when covariances were boosted by classical memory. In practice, these ideal case results can only serve as theoretical benchmarks. Considering then for example the capacity provided by covariances at $M=\infty$ as a benchmark, already a modestly sized ensemble of just $M=10^4$ is enough to rival its IPC when using the univariate CDF, whereas using bivariate CDF achieves more than double this benchmark capacity. Naturally, the advantage grows more significant at larger ensemble sizes. CDF can also be readily used together with classical memory for even greater effect; we present some results regarding this in Appendix~\ref{app:comparison}.

Quantum information processing involving only Gaussian states can be efficiently simulated on a classical computer, and therefore to achieve quantum advantage non-Gaussian states must be used. This also means that non-Gaussian states should outperform Gaussian states, and in order to prove this, all possible performance from Gaussian states should be extracted. While our results do not prove an upper bound for the performance of Gaussian QRC, they show that a significant improvement over the covariance matrix approach is possible.

The first of our results also highlights a difference between DV and CV systems---DV measurement distributions are, as the name implies, discrete, and thus the cumulative distribution function is merely a sum of probabilities of each outcome and cannot provide any additional nonlinearity. On the other hand, for CV systems with non-Gaussian states one might expect the CDF to provide even more nonlinear processing than the moments of the observables.

Application of the classical memory augmented scheme to DV-QRC should presumably be straightforward, though its usefulness should be studied in more detail. There already exist schemes in which the past measurement results are utilized in DV-QRC \cite{kobayashi2024feedback,monomi2025feedback}. However, their approach is quite different in that the measured values are fed back into the reservoir as inputs, rather than being used in training the classical layer. Our results reveal how they could easily be used to boost the capacity at low cost.

Aside from the mentioned non-Gaussian extension and DV-QRC applications, there are further points of interest to research. For example, are there other ways to extract more nonlinearity or processing capacity from Gaussian QRC? How well do the presented methods perform in truly quantum tasks such as state preparation or characterization? We leave these for future work. More generally outside of QRC, it should also be investigated how the CDF can be harnessed as a valuable computational resource also in, e.g., continuous-variable quantum neural networks and what other non-moment based approaches there could be for extracting more computational power out of continuous measurement outcome statistics.

\section{Acknowledgments}
M.H. acknowledges financial support from the Vilho, Yrjö and Kalle Väisälä Foundation and the University of Turku Graduate School. JN gratefully acknowledges financial support from the Academy of Finland under Project No. 348854.


\appendix


\section{Gaussian states and operations} \label{app:gaussian}

The covariance matrix for a single mode Gaussian state is of the form
\begin{equation}
    \sigma = \left(2n_{\text{th}} + 1\right)
    \begin{pmatrix}
        (y+z_c) & z_s\\
        z_s & (y-z_c)
    \end{pmatrix}. 
    \label{eq:covariance}
\end{equation}
Here $n_{\text{th}}$ is the number of thermal excitations, and $y = \cosh(2r)$, $z_s =\sin(\varphi)\sinh(2r)$ and $z_c =\cos(\varphi)\sinh(2r)$ with $r$ and $\varphi$ being the magnitude and phase of squeezing, respectively. The covariance matrix is clearly linear in $n_{\text{th}}$, and nonlinear in $r$ and $\varphi$. In the main text we have chosen to encode the inputs to the reservoir to $r$, but $\varphi$ and $n_{\text{th}}$ are also viable options. Especially when one wishes to consider purely the linear memory capacity of the reservoir, inputs should be encoded to $n_{\text{th}}$.

In the quadrature basis the Hamiltonian can be written in the form 
\begin{equation}
    \mathbf{H} = \frac{1}{2}\mathbf{x}^\top\mathbf{M}\mathbf{x},
\end{equation}
where $\mathbf{x}=\{x_1,p_1,x_2,p_2,\dots\}^\top$ is the vector of quadrature operators and $\mathbf{M}$ is a real symmetric matrix. The corresponding symplectic matrix $\mathbf{S}(t)$ that determines the evolution of the covariance matrices through $\sigma(t) = \mathbf{S}(t) \sigma(0)(\mathbf{S}(t))^\top $ is given by
\begin{equation}
    \mathbf{S}(t) = \exp(\mathbf{\Omega}\mathbf{M}t),
\end{equation}
where $\mathbf{\Omega}_{jk} = -\operatorname{i}[\mathbf{x}_j,\mathbf{x}_k]$ is the symplectic form of commutation relations between the quadrature operators.

Both of the nonlinear crystals, whose Hamiltonians are according to Eq.~\ref{eq:nonlinear_ham}, have corresponding symplectic matrices $S_1(\Delta t)$ and $S_2(\Delta t)$, where $\Delta t$ is the time the pulse spends inside the crystal. The beam splitter is modeled by a symplectic matrix $S_{BS} = \begin{pmatrix}
    \sqrt{R}&\sqrt{1-R}\\
    \sqrt{1-R}&\sqrt{R}
\end{pmatrix}$. 
With these, the total symplectic matrix corresponding to the evolution of the system over one time step is
\begin{equation}
    \mathbf{S}(\Delta t) = \begin{pmatrix}
        \sqrt{R} \mathbf{S}_1(\Delta t)  &\sqrt{1-R} \mathbf{S}_1(\Delta t)\\
        \sqrt{1-R} \mathbf{S}_2(\Delta t)&\sqrt{R} \mathbf{S}_2(\Delta t)
    \end{pmatrix}.
\end{equation}
Considering the covariance matrix of the ancillary modes and the reservoir, which has the general form 
\begin{equation}
    \sigma_k^{\text{(R + anc)}} = \begin{pmatrix}
       \sigma_k^{\text{R}}  & \sigma_k^{\text{corr}} \\ 
        \sigma_k^{\text{corr}}& \sigma_k^{\text{anc}} 
    \end{pmatrix},
\end{equation}
it can be seen that the first column of $\mathbf{S}(\Delta t)$ acts on the reservoir modes, while the second column acts on the ancillary modes. 


\section{Information processing capacity}\label{app:IPC}

Information processing capacity (IPC), introduced in Ref.~\cite{dambre2012information}, quantifies the capability to process past inputs in nonlinear manner. It can be calculated by considering as tasks products of different degree Legendre polynomials of inputs at different delays, 
\begin{equation}
    y_{\{d_i\}} = \prod_{i} \mathcal{P}_{d_i}(s_{k-i}),
\end{equation}
where all possible sets of degrees and delays should be considered. To evaluate the IPC of a system exactly, the sequences should be infinite. However, sufficiently long finite sequences suffice to approximate the IPC with high precision.

Systematically considering all of the possible combinations of degrees and delays requires some care to be taken---preferably tasks which contribute more to the capacity should be calculated first, and when capacities for individual tasks fall below a threshold the process is terminated. We adapt the algorithm used in Ref.~\cite{nokkala2021gaussian} with slight modifications.

For a given degree $d$, a non-decreasing list of delays $\{\tau_1,\tau_2,\dots,\tau_d\}$ is considered. The delays run from $0$ to $\tau_{\mathrm{max}}$, and the lists are ordered such that lists with lower maximum delay are earlier in the order. For example, with $d=3$, the first few lists are $\{0,0,0\}$,$\{0,0,1\}$, $\{0,1,1\}$, $\{1,1,1\}$, $\{0,0,2\}$ $\{0,1,2\}$ \dots. This aims to have the tasks which give higher capacity earlier in the order, so that the procedure can be terminated when a capacity threshold is reached. 

For each list of delays, the target is given by $y_{\{d_i\}} = \prod_{ \tau_i} \mathcal{P}_{m(\tau_i)}(s_{k-i})$, where $m(\tau_i)$ is the multiplicity of $\tau_i$ in the list. For example, if the delay list is $\{1,1,2,3,3\}$, the target is $y = \mathcal{P}_{2}(s_{k-1}) \times \mathcal{P}_{1}(s_{k-2}) \times \mathcal{P}_{2}(s_{k-3})$, with total degree of $5$.

We fix the maximum degree $d_{\mathrm{max}}$ to be $9$, and maximum delay $\tau_{\mathrm{max}}$ to $75$. Then, for each degree up to $d_{\mathrm{max}}$, delay lists are created as described. The targets are calculated, and in the order of the delay lists the capacity of the reservoir to match the target is calculated. Only tasks which contribute more than $10^{-7}$ to the total capacity are counted. When a given number of tasks (100) do not contribute to the capacity, the capacity for that degree is considered to be exhausted and we move on to the next degree. This systematically slightly underestimates the capacity, as it is not guaranteed that the tasks are in order of difficulty: delay list $\{5,5,5\}$ might give higher capacity than $\{1,3,5\}$ due to not requiring cross terms, but it is later in the ordering. However, the effect of this underestimation of IPC should not be too significant especially on a qualitative level.

Since it is known that when using only covariances the reservoir has no capacity to process tasks involving cross terms, the calculation is slightly modified for these cases: only delay lists where all entries are equal are considered. Thus only tasks where the target is a past input or powers thereof are considered.


\section{Gaussian cumulative distribution functions}\label{app:cdf}

The cumulative distribution function $F_X(x)$ of a random variable $X$ is the probability that $X$ is less than or equal to $x$, i.e., $F_X(x) = P(X \leq x)$. If the random variable has a probability density $f_X(x)$, the CDF is given by 
\begin{equation}
    F_X(x) = \int_{-\infty}^x f_X(s) ds.
\end{equation}
A Gaussian random variable $G$ with mean $\mu$ and variance $\sigma^2$ has the probability density 
\begin{equation}
    f_G(x) = \frac{1}{\sqrt{2\pi\sigma^2}} e^{-\frac{(x-\mu)^2}{2\sigma^2}},
\end{equation}
and the CDF
\begin{equation}
    F_G(x) =\int_{-\infty}^x f_G(s) ds = \frac{1}{2} \left[1 +  \operatorname{erf}\left(\frac{x-\mu}{\sigma\sqrt{2}}\right) \right], \label{eq:app-univ_cdf}
\end{equation}
where the error function $\operatorname{erf}(z) = \frac{2}{\sqrt{\pi}}\int_0^z e^{-s^2} ds$ is a special function which cannot be decomposed into elementary functions.  The error function can however be expanded as a Taylor series
\begin{equation}
    \operatorname{erf}(z) 
    = \frac{2}{\sqrt{\pi}}\sum_{n=0}^\infty \frac{(-1)^n z^{2n+1}}{n!(2n+1)} 
    = \frac{2}{\sqrt{\pi}} \left( z - \frac{z^3}{3} + \frac{z^5}{10} - \frac{z^7}{42} + \dots\right). \label{eq:app-erf_taylor}
\end{equation}
When $z = \frac{x-\mu}{\sigma\sqrt{2}}$ is plugged into the above expression, terms containing $\sigma^3$, $\sigma^5$, \dots ($\mu^3$, $\mu^5$, \dots appear. 

The CDF is straightforward to extend to multivariate distributions: for $n$ random variables $\mathbf{X} = (X_1,X_2,\dots, X_n)$ the CDF is defined as $F_{\mathbf{X}}(\mathbf{x}) = P(\mathbf{X}\leq \mathbf{x}) = P(X_1 \leq x_1;X_2 \leq x_2;\dots;X_n \leq x_n)$. With the probability density $f_{\mathbf{X}}(\mathbf{x})$ the CDF is
\begin{equation}
F_{\mathbf{X}}(\mathbf{x}) = \int_{\mathbf{s}<\mathbf{x}}  f_{\mathbf{X}} (\mathbf{s}) d^n \mathbf{s}. \label{eq:app-mvcdf}
\end{equation}
The probability density of a multivariate Gaussian distribution with mean $\mathbf{\bar{x}}$ and covariance matrix $\Sigma$ is
\begin{equation}
    f_\mathbf{G}(\mathbf{x}) = \frac{1}{\sqrt{(2\pi)^n \det(\Sigma)}} e^{-\frac{1}{2}(\mathbf{x}-\bar{\mathbf{x}})^\top \Sigma^{-1} (\mathbf{x}-\bar{\mathbf{x}})},
\end{equation}
which can be inserted into Eq.~\eqref{eq:app-mvcdf} to in theory calculate the CDF. However, this integral cannot be analytically solved even for a bivariate distribution, and numerical methods must be used to evaluate the CDF.

\clearpage
\section{Comparison of the different schemes} \label{app:comparison}

\begin{table}[h]
    \centering
    \begin{tabular}{l|r}
        Scheme & $\mathrm{IPC}/\mathrm{IPC}_\mathrm{cov}$\\
        \hline
        Cov & 1.00\\
        CM1 & 2.00\\
        CM2 & 2.99\\
        UV-CDF10 & 1.74\\
        UV-CDF10 + CM1 & 3.22\\
        UV-CDF10 + CM2 & 4.60\\
        BV-CDF9 & 6.04\\
        BV-CDF9 + CM1 & 11.02\\
        BV-CDF9 + CM2 & 15.23\\
    \end{tabular}
    \caption{Information processing capacities of the different schemes compared to the covariances-only approach. The IPCs are measured as multiples of the covariances-only IPC. \textit{UV-CDF}$n$ (\textit{BV-CDF}$n$) stand for univariate (bivariate) CDF with $n$ sampling points, and CM$P$ stands for classical memory of $P$ past time steps. Averages over 50 realizations.}
    \label{tab:app-ipc-table}
\end{table}

Here we briefly compare the IPC of the different schemes to the covariances-only approach. For completeness we include the case where both the CDF results and their delayed versions are included, such that the observable vector at each time step is $\mathbf{O}_k = \mathbf{o}_k \oplus \mathbf{o}_{k-1}\oplus\dots$, with each $\mathbf{o}_k$ containing both the entries of the covariance matrix and the CDF results. Table~\ref{tab:app-ipc-table} shows the comparison of IPCs in the ideal case of $M \to \infty$, highlighting the compatibility of the two methods introduced in this article: when the bivariate CDF is used in conjunction with classical memory, the capacities are well over ten times that of the covariance-only approach.


\bibliography{bibliography}

\begin{thebibliography}{22}%
\makeatletter
\providecommand \@ifxundefined [1]{%
 \@ifx{#1\undefined}
}%
\providecommand \@ifnum [1]{%
 \ifnum #1\expandafter \@firstoftwo
 \else \expandafter \@secondoftwo
 \fi
}%
\providecommand \@ifx [1]{%
 \ifx #1\expandafter \@firstoftwo
 \else \expandafter \@secondoftwo
 \fi
}%
\providecommand \natexlab [1]{#1}%
\providecommand \enquote  [1]{``#1''}%
\providecommand \bibnamefont  [1]{#1}%
\providecommand \bibfnamefont [1]{#1}%
\providecommand \citenamefont [1]{#1}%
\providecommand \href@noop [0]{\@secondoftwo}%
\providecommand \href [0]{\begingroup \@sanitize@url \@href}%
\providecommand \@href[1]{\@@startlink{#1}\@@href}%
\providecommand \@@href[1]{\endgroup#1\@@endlink}%
\providecommand \@sanitize@url [0]{\catcode `\\12\catcode `\$12\catcode `\&12\catcode `\#12\catcode `\^12\catcode `\_12\catcode `\%12\relax}%
\providecommand \@@startlink[1]{}%
\providecommand \@@endlink[0]{}%
\providecommand \url  [0]{\begingroup\@sanitize@url \@url }%
\providecommand \@url [1]{\endgroup\@href {#1}{\urlprefix }}%
\providecommand \urlprefix  [0]{URL }%
\providecommand \Eprint [0]{\href }%
\providecommand \doibase [0]{https://doi.org/}%
\providecommand \selectlanguage [0]{\@gobble}%
\providecommand \bibinfo  [0]{\@secondoftwo}%
\providecommand \bibfield  [0]{\@secondoftwo}%
\providecommand \translation [1]{[#1]}%
\providecommand \BibitemOpen [0]{}%
\providecommand \bibitemStop [0]{}%
\providecommand \bibitemNoStop [0]{.\EOS\space}%
\providecommand \EOS [0]{\spacefactor3000\relax}%
\providecommand \BibitemShut  [1]{\csname bibitem#1\endcsname}%
\let\auto@bib@innerbib\@empty
\bibitem [{\citenamefont {Mujal}\ \emph {et~al.}(2021)\citenamefont {Mujal}, \citenamefont {Mart{\'\i}nez-Pe{\~n}a}, \citenamefont {Nokkala}, \citenamefont {Garc{\'\i}a-Beni}, \citenamefont {Giorgi}, \citenamefont {Soriano},\ and\ \citenamefont {Zambrini}}]{mujal2021opportunities}%
  \BibitemOpen
  \bibfield  {author} {\bibinfo {author} {\bibfnamefont {P.}~\bibnamefont {Mujal}}, \bibinfo {author} {\bibfnamefont {R.}~\bibnamefont {Mart{\'\i}nez-Pe{\~n}a}}, \bibinfo {author} {\bibfnamefont {J.}~\bibnamefont {Nokkala}}, \bibinfo {author} {\bibfnamefont {J.}~\bibnamefont {Garc{\'\i}a-Beni}}, \bibinfo {author} {\bibfnamefont {G.~L.}\ \bibnamefont {Giorgi}}, \bibinfo {author} {\bibfnamefont {M.~C.}\ \bibnamefont {Soriano}},\ and\ \bibinfo {author} {\bibfnamefont {R.}~\bibnamefont {Zambrini}},\ }\href@noop {} {\bibfield  {journal} {\bibinfo  {journal} {Advanced Quantum Technologies}\ }\textbf {\bibinfo {volume} {4}},\ \bibinfo {pages} {2100027} (\bibinfo {year} {2021})}\BibitemShut {NoStop}%
\bibitem [{\citenamefont {Fujii}\ and\ \citenamefont {Nakajima}(2017)}]{fujii2017harnessing}%
  \BibitemOpen
  \bibfield  {author} {\bibinfo {author} {\bibfnamefont {K.}~\bibnamefont {Fujii}}\ and\ \bibinfo {author} {\bibfnamefont {K.}~\bibnamefont {Nakajima}},\ }\href@noop {} {\bibfield  {journal} {\bibinfo  {journal} {Physical Review Applied}\ }\textbf {\bibinfo {volume} {8}},\ \bibinfo {pages} {024030} (\bibinfo {year} {2017})}\BibitemShut {NoStop}%
\bibitem [{\citenamefont {Ghosh}\ \emph {et~al.}(2020)\citenamefont {Ghosh}, \citenamefont {Opala}, \citenamefont {Matuszewski}, \citenamefont {Paterek},\ and\ \citenamefont {Liew}}]{ghosh2020reconstructing}%
  \BibitemOpen
  \bibfield  {author} {\bibinfo {author} {\bibfnamefont {S.}~\bibnamefont {Ghosh}}, \bibinfo {author} {\bibfnamefont {A.}~\bibnamefont {Opala}}, \bibinfo {author} {\bibfnamefont {M.}~\bibnamefont {Matuszewski}}, \bibinfo {author} {\bibfnamefont {T.}~\bibnamefont {Paterek}},\ and\ \bibinfo {author} {\bibfnamefont {T.~C.}\ \bibnamefont {Liew}},\ }\href@noop {} {\bibfield  {journal} {\bibinfo  {journal} {IEEE Transactions on Neural Networks and Learning Systems}\ }\textbf {\bibinfo {volume} {32}},\ \bibinfo {pages} {3148} (\bibinfo {year} {2020})}\BibitemShut {NoStop}%
\bibitem [{\citenamefont {Senanian}\ \emph {et~al.}(2024)\citenamefont {Senanian}, \citenamefont {Prabhu}, \citenamefont {Kremenetski}, \citenamefont {Roy}, \citenamefont {Cao}, \citenamefont {Kline}, \citenamefont {Onodera}, \citenamefont {Wright}, \citenamefont {Wu}, \citenamefont {Fatemi},\ and\ \citenamefont {McMahon}}]{senanian2024microwave}%
  \BibitemOpen
  \bibfield  {author} {\bibinfo {author} {\bibfnamefont {A.}~\bibnamefont {Senanian}}, \bibinfo {author} {\bibfnamefont {S.}~\bibnamefont {Prabhu}}, \bibinfo {author} {\bibfnamefont {V.}~\bibnamefont {Kremenetski}}, \bibinfo {author} {\bibfnamefont {S.}~\bibnamefont {Roy}}, \bibinfo {author} {\bibfnamefont {Y.}~\bibnamefont {Cao}}, \bibinfo {author} {\bibfnamefont {J.}~\bibnamefont {Kline}}, \bibinfo {author} {\bibfnamefont {T.}~\bibnamefont {Onodera}}, \bibinfo {author} {\bibfnamefont {L.~G.}\ \bibnamefont {Wright}}, \bibinfo {author} {\bibfnamefont {X.}~\bibnamefont {Wu}}, \bibinfo {author} {\bibfnamefont {V.}~\bibnamefont {Fatemi}},\ and\ \bibinfo {author} {\bibfnamefont {P.~L.}\ \bibnamefont {McMahon}},\ }\href@noop {} {\bibfield  {journal} {\bibinfo  {journal} {Nature Communications}\ }\textbf {\bibinfo {volume} {15}},\ \bibinfo {pages} {7490} (\bibinfo {year} {2024})}\BibitemShut {NoStop}%
\bibitem [{\citenamefont {Krisnanda}\ \emph {et~al.}(2025)\citenamefont {Krisnanda}, \citenamefont {Song}, \citenamefont {Copetudo}, \citenamefont {Fontaine}, \citenamefont {Paterek}, \citenamefont {Liew},\ and\ \citenamefont {Gao}}]{krisnanda2025experimental}%
  \BibitemOpen
  \bibfield  {author} {\bibinfo {author} {\bibfnamefont {T.}~\bibnamefont {Krisnanda}}, \bibinfo {author} {\bibfnamefont {P.}~\bibnamefont {Song}}, \bibinfo {author} {\bibfnamefont {A.}~\bibnamefont {Copetudo}}, \bibinfo {author} {\bibfnamefont {C.~Y.}\ \bibnamefont {Fontaine}}, \bibinfo {author} {\bibfnamefont {T.}~\bibnamefont {Paterek}}, \bibinfo {author} {\bibfnamefont {T.~C.~H.}\ \bibnamefont {Liew}},\ and\ \bibinfo {author} {\bibfnamefont {Y.~Y.}\ \bibnamefont {Gao}},\ }\href {https://doi.org/10.1088/2058-9565/addffe} {\bibfield  {journal} {\bibinfo  {journal} {Quantum Science and Technology}\ }\textbf {\bibinfo {volume} {10}},\ \bibinfo {pages} {035041} (\bibinfo {year} {2025})}\BibitemShut {NoStop}%
\bibitem [{\citenamefont {Mujal}\ \emph {et~al.}(2023)\citenamefont {Mujal}, \citenamefont {Mart{\'\i}nez-Pe{\~n}a}, \citenamefont {Giorgi}, \citenamefont {Soriano},\ and\ \citenamefont {Zambrini}}]{mujal2023time}%
  \BibitemOpen
  \bibfield  {author} {\bibinfo {author} {\bibfnamefont {P.}~\bibnamefont {Mujal}}, \bibinfo {author} {\bibfnamefont {R.}~\bibnamefont {Mart{\'\i}nez-Pe{\~n}a}}, \bibinfo {author} {\bibfnamefont {G.~L.}\ \bibnamefont {Giorgi}}, \bibinfo {author} {\bibfnamefont {M.~C.}\ \bibnamefont {Soriano}},\ and\ \bibinfo {author} {\bibfnamefont {R.}~\bibnamefont {Zambrini}},\ }\href@noop {} {\bibfield  {journal} {\bibinfo  {journal} {npj Quantum Information}\ }\textbf {\bibinfo {volume} {9}},\ \bibinfo {pages} {16} (\bibinfo {year} {2023})}\BibitemShut {NoStop}%
\bibitem [{\citenamefont {Kobayashi}\ \emph {et~al.}(2024)\citenamefont {Kobayashi}, \citenamefont {Fujii},\ and\ \citenamefont {Yamamoto}}]{kobayashi2024feedback}%
  \BibitemOpen
  \bibfield  {author} {\bibinfo {author} {\bibfnamefont {K.}~\bibnamefont {Kobayashi}}, \bibinfo {author} {\bibfnamefont {K.}~\bibnamefont {Fujii}},\ and\ \bibinfo {author} {\bibfnamefont {N.}~\bibnamefont {Yamamoto}},\ }\href@noop {} {\bibfield  {journal} {\bibinfo  {journal} {PRX Quantum}\ }\textbf {\bibinfo {volume} {5}},\ \bibinfo {pages} {040325} (\bibinfo {year} {2024})}\BibitemShut {NoStop}%
\bibitem [{\citenamefont {Monomi}\ \emph {et~al.}(2025)\citenamefont {Monomi}, \citenamefont {Setoyama},\ and\ \citenamefont {Hasegawa}}]{monomi2025feedback}%
  \BibitemOpen
  \bibfield  {author} {\bibinfo {author} {\bibfnamefont {T.}~\bibnamefont {Monomi}}, \bibinfo {author} {\bibfnamefont {W.}~\bibnamefont {Setoyama}},\ and\ \bibinfo {author} {\bibfnamefont {Y.}~\bibnamefont {Hasegawa}},\ }\href {https://arxiv.org/abs/2503.17939} {\bibinfo {title} {Feedback-enhanced quantum reservoir computing with weak measurements}} (\bibinfo {year} {2025}),\ \Eprint {https://arxiv.org/abs/2503.17939} {arXiv:2503.17939 [quant-ph]} \BibitemShut {NoStop}%
\bibitem [{\citenamefont {Paparelle}\ \emph {et~al.}(2025)\citenamefont {Paparelle}, \citenamefont {Henaff}, \citenamefont {Garcia-Beni}, \citenamefont {Gillet}, \citenamefont {Giorgi}, \citenamefont {Soriano}, \citenamefont {Zambrini},\ and\ \citenamefont {Parigi}}]{paparelle2025experimental}%
  \BibitemOpen
  \bibfield  {author} {\bibinfo {author} {\bibfnamefont {I.}~\bibnamefont {Paparelle}}, \bibinfo {author} {\bibfnamefont {J.}~\bibnamefont {Henaff}}, \bibinfo {author} {\bibfnamefont {J.}~\bibnamefont {Garcia-Beni}}, \bibinfo {author} {\bibfnamefont {E.}~\bibnamefont {Gillet}}, \bibinfo {author} {\bibfnamefont {G.~L.}\ \bibnamefont {Giorgi}}, \bibinfo {author} {\bibfnamefont {M.~C.}\ \bibnamefont {Soriano}}, \bibinfo {author} {\bibfnamefont {R.}~\bibnamefont {Zambrini}},\ and\ \bibinfo {author} {\bibfnamefont {V.}~\bibnamefont {Parigi}},\ }\href {https://arxiv.org/abs/2506.07279} {\bibinfo {title} {Experimental memory control in continuous variable optical quantum reservoir computing}} (\bibinfo {year} {2025}),\ \Eprint {https://arxiv.org/abs/2506.07279} {arXiv:2506.07279 [quant-ph]} \BibitemShut {NoStop}%
\bibitem [{\citenamefont {Garc{\'\i}a-Beni}\ \emph {et~al.}(2023)\citenamefont {Garc{\'\i}a-Beni}, \citenamefont {Giorgi}, \citenamefont {Soriano},\ and\ \citenamefont {Zambrini}}]{garcia2023scalable}%
  \BibitemOpen
  \bibfield  {author} {\bibinfo {author} {\bibfnamefont {J.}~\bibnamefont {Garc{\'\i}a-Beni}}, \bibinfo {author} {\bibfnamefont {G.~L.}\ \bibnamefont {Giorgi}}, \bibinfo {author} {\bibfnamefont {M.~C.}\ \bibnamefont {Soriano}},\ and\ \bibinfo {author} {\bibfnamefont {R.}~\bibnamefont {Zambrini}},\ }\href@noop {} {\bibfield  {journal} {\bibinfo  {journal} {Physical Review Applied}\ }\textbf {\bibinfo {volume} {20}},\ \bibinfo {pages} {014051} (\bibinfo {year} {2023})}\BibitemShut {NoStop}%
\bibitem [{\citenamefont {Nokkala}\ \emph {et~al.}(2021)\citenamefont {Nokkala}, \citenamefont {Mart{\'\i}nez-Pe{\~n}a}, \citenamefont {Giorgi}, \citenamefont {Parigi}, \citenamefont {Soriano},\ and\ \citenamefont {Zambrini}}]{nokkala2021gaussian}%
  \BibitemOpen
  \bibfield  {author} {\bibinfo {author} {\bibfnamefont {J.}~\bibnamefont {Nokkala}}, \bibinfo {author} {\bibfnamefont {R.}~\bibnamefont {Mart{\'\i}nez-Pe{\~n}a}}, \bibinfo {author} {\bibfnamefont {G.~L.}\ \bibnamefont {Giorgi}}, \bibinfo {author} {\bibfnamefont {V.}~\bibnamefont {Parigi}}, \bibinfo {author} {\bibfnamefont {M.~C.}\ \bibnamefont {Soriano}},\ and\ \bibinfo {author} {\bibfnamefont {R.}~\bibnamefont {Zambrini}},\ }\href@noop {} {\bibfield  {journal} {\bibinfo  {journal} {Communications Physics}\ }\textbf {\bibinfo {volume} {4}},\ \bibinfo {pages} {53} (\bibinfo {year} {2021})}\BibitemShut {NoStop}%
\bibitem [{\citenamefont {Nokkala}\ \emph {et~al.}(2022)\citenamefont {Nokkala}, \citenamefont {Mart{\'\i}nez-Pe{\~n}a}, \citenamefont {Zambrini},\ and\ \citenamefont {Soriano}}]{nokkala2022high}%
  \BibitemOpen
  \bibfield  {author} {\bibinfo {author} {\bibfnamefont {J.}~\bibnamefont {Nokkala}}, \bibinfo {author} {\bibfnamefont {R.}~\bibnamefont {Mart{\'\i}nez-Pe{\~n}a}}, \bibinfo {author} {\bibfnamefont {R.}~\bibnamefont {Zambrini}},\ and\ \bibinfo {author} {\bibfnamefont {M.~C.}\ \bibnamefont {Soriano}},\ }\href@noop {} {\bibfield  {journal} {\bibinfo  {journal} {IEEE Transactions on Neural Networks and Learning Systems}\ }\textbf {\bibinfo {volume} {33}},\ \bibinfo {pages} {2664} (\bibinfo {year} {2022})}\BibitemShut {NoStop}%
\bibitem [{\citenamefont {Nokkala}\ \emph {et~al.}(2024)\citenamefont {Nokkala}, \citenamefont {Giorgi},\ and\ \citenamefont {Zambrini}}]{nokkala2024retrieving}%
  \BibitemOpen
  \bibfield  {author} {\bibinfo {author} {\bibfnamefont {J.}~\bibnamefont {Nokkala}}, \bibinfo {author} {\bibfnamefont {G.~L.}\ \bibnamefont {Giorgi}},\ and\ \bibinfo {author} {\bibfnamefont {R.}~\bibnamefont {Zambrini}},\ }\href {https://doi.org/10.1088/2632-2153/ad5f12} {\bibfield  {journal} {\bibinfo  {journal} {Machine Learning: Science and Technology}\ }\textbf {\bibinfo {volume} {5}},\ \bibinfo {pages} {035022} (\bibinfo {year} {2024})}\BibitemShut {NoStop}%
\bibitem [{\citenamefont {Garc{\'\i}a-Beni}\ \emph {et~al.}(2024)\citenamefont {Garc{\'\i}a-Beni}, \citenamefont {Luca~Giorgi}, \citenamefont {Soriano},\ and\ \citenamefont {Zambrini}}]{garcia2024squeezing}%
  \BibitemOpen
  \bibfield  {author} {\bibinfo {author} {\bibfnamefont {J.}~\bibnamefont {Garc{\'\i}a-Beni}}, \bibinfo {author} {\bibfnamefont {G.}~\bibnamefont {Luca~Giorgi}}, \bibinfo {author} {\bibfnamefont {M.~C.}\ \bibnamefont {Soriano}},\ and\ \bibinfo {author} {\bibfnamefont {R.}~\bibnamefont {Zambrini}},\ }\href@noop {} {\bibfield  {journal} {\bibinfo  {journal} {Optics Express}\ }\textbf {\bibinfo {volume} {32}},\ \bibinfo {pages} {6733} (\bibinfo {year} {2024})}\BibitemShut {NoStop}%
\bibitem [{\citenamefont {Sakemi}\ \emph {et~al.}(2020)\citenamefont {Sakemi}, \citenamefont {Morino}, \citenamefont {Leleu},\ and\ \citenamefont {Aihara}}]{sakemi2020model}%
  \BibitemOpen
  \bibfield  {author} {\bibinfo {author} {\bibfnamefont {Y.}~\bibnamefont {Sakemi}}, \bibinfo {author} {\bibfnamefont {K.}~\bibnamefont {Morino}}, \bibinfo {author} {\bibfnamefont {T.}~\bibnamefont {Leleu}},\ and\ \bibinfo {author} {\bibfnamefont {K.}~\bibnamefont {Aihara}},\ }\href@noop {} {\bibfield  {journal} {\bibinfo  {journal} {Scientific reports}\ }\textbf {\bibinfo {volume} {10}},\ \bibinfo {pages} {21794} (\bibinfo {year} {2020})}\BibitemShut {NoStop}%
\bibitem [{\citenamefont {Duan}\ \emph {et~al.}(2023)\citenamefont {Duan}, \citenamefont {Ying}, \citenamefont {Leng}, \citenamefont {Kurths}, \citenamefont {Lin},\ and\ \citenamefont {Ma}}]{duan2023embedding}%
  \BibitemOpen
  \bibfield  {author} {\bibinfo {author} {\bibfnamefont {X.-Y.}\ \bibnamefont {Duan}}, \bibinfo {author} {\bibfnamefont {X.}~\bibnamefont {Ying}}, \bibinfo {author} {\bibfnamefont {S.-Y.}\ \bibnamefont {Leng}}, \bibinfo {author} {\bibfnamefont {J.}~\bibnamefont {Kurths}}, \bibinfo {author} {\bibfnamefont {W.}~\bibnamefont {Lin}},\ and\ \bibinfo {author} {\bibfnamefont {H.-F.}\ \bibnamefont {Ma}},\ }\href@noop {} {\bibfield  {journal} {\bibinfo  {journal} {Physical Review Research}\ }\textbf {\bibinfo {volume} {5}},\ \bibinfo {pages} {L022041} (\bibinfo {year} {2023})}\BibitemShut {NoStop}%
\bibitem [{\citenamefont {Ahmed}\ \emph {et~al.}(2025)\citenamefont {Ahmed}, \citenamefont {Tennie},\ and\ \citenamefont {Magri}}]{ahmed2025optimal}%
  \BibitemOpen
  \bibfield  {author} {\bibinfo {author} {\bibfnamefont {O.}~\bibnamefont {Ahmed}}, \bibinfo {author} {\bibfnamefont {F.}~\bibnamefont {Tennie}},\ and\ \bibinfo {author} {\bibfnamefont {L.}~\bibnamefont {Magri}},\ }\href@noop {} {\bibfield  {journal} {\bibinfo  {journal} {Quantum Machine Intelligence}\ }\textbf {\bibinfo {volume} {7}},\ \bibinfo {pages} {1} (\bibinfo {year} {2025})}\BibitemShut {NoStop}%
\bibitem [{\citenamefont {Weedbrook}\ \emph {et~al.}(2012)\citenamefont {Weedbrook}, \citenamefont {Pirandola}, \citenamefont {Garc{\'\i}a-Patr{\'o}n}, \citenamefont {Cerf}, \citenamefont {Ralph}, \citenamefont {Shapiro},\ and\ \citenamefont {Lloyd}}]{weedbrook2012gaussian}%
  \BibitemOpen
  \bibfield  {author} {\bibinfo {author} {\bibfnamefont {C.}~\bibnamefont {Weedbrook}}, \bibinfo {author} {\bibfnamefont {S.}~\bibnamefont {Pirandola}}, \bibinfo {author} {\bibfnamefont {R.}~\bibnamefont {Garc{\'\i}a-Patr{\'o}n}}, \bibinfo {author} {\bibfnamefont {N.~J.}\ \bibnamefont {Cerf}}, \bibinfo {author} {\bibfnamefont {T.~C.}\ \bibnamefont {Ralph}}, \bibinfo {author} {\bibfnamefont {J.~H.}\ \bibnamefont {Shapiro}},\ and\ \bibinfo {author} {\bibfnamefont {S.}~\bibnamefont {Lloyd}},\ }\href@noop {} {\bibfield  {journal} {\bibinfo  {journal} {Reviews of Modern Physics}\ }\textbf {\bibinfo {volume} {84}},\ \bibinfo {pages} {621} (\bibinfo {year} {2012})}\BibitemShut {NoStop}%
\bibitem [{\citenamefont {Wishart}(1928)}]{wishart1928generalised}%
  \BibitemOpen
  \bibfield  {author} {\bibinfo {author} {\bibfnamefont {J.}~\bibnamefont {Wishart}},\ }\href@noop {} {\bibfield  {journal} {\bibinfo  {journal} {Biometrika}\ ,\ \bibinfo {pages} {32}} (\bibinfo {year} {1928})}\BibitemShut {NoStop}%
\bibitem [{\citenamefont {Kubota}\ \emph {et~al.}(2021)\citenamefont {Kubota}, \citenamefont {Takahashi},\ and\ \citenamefont {Nakajima}}]{kubota2021unifying}%
  \BibitemOpen
  \bibfield  {author} {\bibinfo {author} {\bibfnamefont {T.}~\bibnamefont {Kubota}}, \bibinfo {author} {\bibfnamefont {H.}~\bibnamefont {Takahashi}},\ and\ \bibinfo {author} {\bibfnamefont {K.}~\bibnamefont {Nakajima}},\ }\href@noop {} {\bibfield  {journal} {\bibinfo  {journal} {Physical Review Research}\ }\textbf {\bibinfo {volume} {3}},\ \bibinfo {pages} {043135} (\bibinfo {year} {2021})}\BibitemShut {NoStop}%
\bibitem [{\citenamefont {Dambre}\ \emph {et~al.}(2012)\citenamefont {Dambre}, \citenamefont {Verstraeten}, \citenamefont {Schrauwen},\ and\ \citenamefont {Massar}}]{dambre2012information}%
  \BibitemOpen
  \bibfield  {author} {\bibinfo {author} {\bibfnamefont {J.}~\bibnamefont {Dambre}}, \bibinfo {author} {\bibfnamefont {D.}~\bibnamefont {Verstraeten}}, \bibinfo {author} {\bibfnamefont {B.}~\bibnamefont {Schrauwen}},\ and\ \bibinfo {author} {\bibfnamefont {S.}~\bibnamefont {Massar}},\ }\href@noop {} {\bibfield  {journal} {\bibinfo  {journal} {Scientific Reports}\ }\textbf {\bibinfo {volume} {2}},\ \bibinfo {pages} {514} (\bibinfo {year} {2012})}\BibitemShut {NoStop}%
\bibitem [{\citenamefont {Virtanen}\ \emph {et~al.}(2020)\citenamefont {Virtanen}, \citenamefont {Gommers}, \citenamefont {Oliphant}, \citenamefont {Haberland}, \citenamefont {Reddy}, \citenamefont {Cournapeau}, \citenamefont {Burovski}, \citenamefont {Peterson}, \citenamefont {Weckesser}, \citenamefont {Bright}, \citenamefont {{van der Walt}}, \citenamefont {Brett}, \citenamefont {Wilson}, \citenamefont {Millman}, \citenamefont {Mayorov}, \citenamefont {Nelson}, \citenamefont {Jones}, \citenamefont {Kern}, \citenamefont {Larson}, \citenamefont {Carey}, \citenamefont {Polat}, \citenamefont {Feng}, \citenamefont {Moore}, \citenamefont {{VanderPlas}}, \citenamefont {Laxalde}, \citenamefont {Perktold}, \citenamefont {Cimrman}, \citenamefont {Henriksen}, \citenamefont {Quintero}, \citenamefont {Harris}, \citenamefont {Archibald}, \citenamefont {Ribeiro}, \citenamefont {Pedregosa}, \citenamefont {{van Mulbregt}},\ and\ \citenamefont {{SciPy 1.0 Contributors}}}]{virtanen2020scipy}%
  \BibitemOpen
  \bibfield  {author} {\bibinfo {author} {\bibfnamefont {P.}~\bibnamefont {Virtanen}}, \bibinfo {author} {\bibfnamefont {R.}~\bibnamefont {Gommers}}, \bibinfo {author} {\bibfnamefont {T.~E.}\ \bibnamefont {Oliphant}}, \bibinfo {author} {\bibfnamefont {M.}~\bibnamefont {Haberland}}, \bibinfo {author} {\bibfnamefont {T.}~\bibnamefont {Reddy}}, \bibinfo {author} {\bibfnamefont {D.}~\bibnamefont {Cournapeau}}, \bibinfo {author} {\bibfnamefont {E.}~\bibnamefont {Burovski}}, \bibinfo {author} {\bibfnamefont {P.}~\bibnamefont {Peterson}}, \bibinfo {author} {\bibfnamefont {W.}~\bibnamefont {Weckesser}}, \bibinfo {author} {\bibfnamefont {J.}~\bibnamefont {Bright}}, \bibinfo {author} {\bibfnamefont {S.~J.}\ \bibnamefont {{van der Walt}}}, \bibinfo {author} {\bibfnamefont {M.}~\bibnamefont {Brett}}, \bibinfo {author} {\bibfnamefont {J.}~\bibnamefont {Wilson}}, \bibinfo {author} {\bibfnamefont {K.~J.}\ \bibnamefont {Millman}}, \bibinfo {author} {\bibfnamefont {N.}~\bibnamefont {Mayorov}}, \bibinfo {author} {\bibfnamefont
  {A.~R.~J.}\ \bibnamefont {Nelson}}, \bibinfo {author} {\bibfnamefont {E.}~\bibnamefont {Jones}}, \bibinfo {author} {\bibfnamefont {R.}~\bibnamefont {Kern}}, \bibinfo {author} {\bibfnamefont {E.}~\bibnamefont {Larson}}, \bibinfo {author} {\bibfnamefont {C.~J.}\ \bibnamefont {Carey}}, \bibinfo {author} {\bibfnamefont {{\.I}.}~\bibnamefont {Polat}}, \bibinfo {author} {\bibfnamefont {Y.}~\bibnamefont {Feng}}, \bibinfo {author} {\bibfnamefont {E.~W.}\ \bibnamefont {Moore}}, \bibinfo {author} {\bibfnamefont {J.}~\bibnamefont {{VanderPlas}}}, \bibinfo {author} {\bibfnamefont {D.}~\bibnamefont {Laxalde}}, \bibinfo {author} {\bibfnamefont {J.}~\bibnamefont {Perktold}}, \bibinfo {author} {\bibfnamefont {R.}~\bibnamefont {Cimrman}}, \bibinfo {author} {\bibfnamefont {I.}~\bibnamefont {Henriksen}}, \bibinfo {author} {\bibfnamefont {E.~A.}\ \bibnamefont {Quintero}}, \bibinfo {author} {\bibfnamefont {C.~R.}\ \bibnamefont {Harris}}, \bibinfo {author} {\bibfnamefont {A.~M.}\ \bibnamefont {Archibald}}, \bibinfo {author}
  {\bibfnamefont {A.~H.}\ \bibnamefont {Ribeiro}}, \bibinfo {author} {\bibfnamefont {F.}~\bibnamefont {Pedregosa}}, \bibinfo {author} {\bibfnamefont {P.}~\bibnamefont {{van Mulbregt}}},\ and\ \bibinfo {author} {\bibnamefont {{SciPy 1.0 Contributors}}},\ }\href {https://doi.org/10.1038/s41592-019-0686-2} {\bibfield  {journal} {\bibinfo  {journal} {Nature Methods}\ }\textbf {\bibinfo {volume} {17}},\ \bibinfo {pages} {261} (\bibinfo {year} {2020})}\BibitemShut {NoStop}%
\end{thebibliography}%

\end{document}